\documentclass[prc,aps,floatfix,groupedaddress,amsmath,amssymb]{revtex4}
\usepackage{graphicx}
\usepackage{epsfig,ulem}
\usepackage{dcolumn}
\usepackage{bm}

\def\beq{\begin{equation}}
\def\eeq{\end{equation}}

\newcommand{\rv}{\mathbf{r}}
\newcommand{\rj}{\mathbf{j}}
\newcommand{\kv}{\mathbf{k}}
\newcommand{\qv}{\mathbf{q}}

\newcommand{\kd}{{\rm k}}

\newcommand{\mbG}{\mathbf{G}}

\newcommand{\mV}{\mathcal{V}}

\newcommand{\be}{\begin{eqnarray}}
\newcommand{\ee}{\end{eqnarray}}
\newcommand{\nn}{\nonumber}
\newcommand{\een}{\nonumber\end{eqnarray}}

\begin{document}
\thispagestyle{empty}
\vspace*{0.5 cm}
\begin{center}
{\bf Plasmons in strong superconductors}
\\
\vspace*{1cm} {\bf M. Baldo$^1$ and C. Ducoin}$^2$\\
\vspace*{.3cm}
$^1${\it INFN, Sezione di Catania,
 Via S. Sofia 64, I-95123, Catania, Italy} \\
\vspace*{.3cm} $^2${\it CFC, Department of Physics, University of Coimbra, P3004-516, Coimbra, Portugal.}
\vspace*{1 cm}
\end{center}
{\bf ABSTRACT} \\
We present a study of the possible plasmon excitations that can occur in systems 
where strong superconductivity is present. 
In these systems the plasmon energy is comparable to,
or smaller than the pairing gap. 
As a prototype of these systems we consider the proton component of neutron star matter 
just below the crust when electron screening is 
not taken into account. 
For the realistic case we consider in detail the different aspects of the elementary excitations
when the  proton, electron components are considered within the Random Phase Approximation generalized to the
superfluid case, while the influence of the neutron component is considered only at qualitative level. Electron
screening plays a major role in modifying the proton spectrum and spectral function. At the same time the electron
plasmon is strongly modified and damped by the indirect coupling with the superfluid proton component, even at
moderately low values of the gap. The excitation spectrum shows the interplay of the different components and
their relevance for each excitation mode. 
The results are relevant for neutrino physics and thermodynamical
processes in neutron stars. If electron screening is neglected, the spectral properties of the proton component
show some resemblance with the physical situation in high T$_c$ superconductors, and we briefly discuss
similarities and differences in this connection. In a general prospect, the results of the study emphasize the
role of Coulomb interaction in strong superconductors.

 \vskip 0.3 cm
PACS :
21.65.+f ,  
24.10.Cn ,  
26.60.+c ,  
03.75.Ss    


\section{Introduction}

The excitation spectrum in superconductors is strongly affected by the long range Coulomb interaction. In the
simplest case of superconductivity in metals, the long range Coulomb interaction is responsible of the "Higgs
mechanism" that produces a non-zero energy excitation at zero momentum. 
The otherwise present Goldstone mode, caused by
the breaking of gauge invariance in a superconductor, actually disappears.
Instead, a plasmon-like mode appears well above the gap energy, mainly at the plasmon energy of the normal metal~\cite{Schriefferpr,Schriefferb}.
In this way, the sound-like spectrum, below twice the energy gap, typical of e.g.
neutral s-wave pairing superfluids, is modified to a spectrum that starts at finite high energy at zero momentum
and varies smoothly with momentum.

More complex is the situation in a high T$_c$ superconductor, since electrons have no Fermi liquid behavior.
However, it has been argued~\cite{hightc1} that along the so called c-axis plasmon-like excitations ("Josephson
plasmons") should be present, and indeed some authors have claimed that they can be experimentally observed~\cite{hightc2,hightc3}. 
In this case, the plasmon frequency is smaller than the energy gap.

The system over which we will focus our study is the homogeneous matter in neutron stars, just below the crust,
that is expected to be superfluid. In particular,
 homogeneous matter below Neutron Star (NS) crust is expected to have a proton superfluid component. The
elementary excitations of the matter affect the whole thermodynamics and long term evolution of the star. Since
the main components of the matter are neutrons, protons, electrons and muons~\cite{shap}, the spectral properties
of these excitations can have a complex structure.  Collective modes in asymmetric nuclear matter have been
studied previously, e.g. in Refs.~\cite{Haensel-NPA301, Matera-PRC49, Greco-PRC67}. In the astrophysical context,
a study of the collective excitations in normal neutron star matter on the basis of the relativistic mean field
method has been presented in Ref.~\cite{Providencia-PRC74}. The spectral functions of the different components in
normal neutron star matter have been calculated in Ref.~\cite{paper1} on the basis of non-relativistic
Random-Phase-Approximation (RPA) for the nucleonic components and relativistic RPA for the leptonic components.
Different models for the nuclear effective interaction have been considered and a detailed comparison was done
between some Skyrme forces and a microscopically derived interaction. 

In Ref.~\cite{paper2} particular attention was payed to the plasmon excitations of the proton component, where
the Coulomb interaction, among protons and between protons and electrons, plays a major role.
In this work we will extend this study to the case of superfluid matter. The elementary excitations in superfluid
neutron star matter have been studied by several authors, often with controversial results
~\cite{Reddy,Kundu,Leinson1,Armen,Leinson2,Vosk}. It is well known that a neutral superfluid must present a
Goldstone mode at low momentum, and this can have a strong influence on e.g. neutrino
emission~\cite{Yako,Reddy,Kundu,Leinson1,Armen,Leinson2,Vosk} or mean free path. As already mentioned, for a
single charged superfluid the Coulomb interaction suppresses the mode, which is replaced by the plasmon mode. In
neutron star matter the physical situation is complicated by the multi-component structure, since the plasmon mode
is in fact mainly an electron excitation, and furthermore the nuclear interaction couples neutron and proton
excitation modes. We formulate the general theoretical scheme within the conserving approximations~\cite{BaymKad,Baym}, 
which guarantee current conservation and the fulfilment of the related Generalized Ward
Identities (GWI). However, we will focus the study only on the 
charged components, protons and electrons, 
while the complete treatment will be presented elsewhere in future works. 
The reason of this choice is that the role of Coulomb interaction is
quite crucial in this case, while the coupling with neutron is expected to be weak~\cite{paper1}. 
What is left out from the treatment is the possible role of the entrainment~\cite{chamel,haensel} 
when both neutron and proton components are simultaneously superfluid. 
However, below the crust neutron superfluidity disappears rapidly at increasing density, 
and we will therefore assume that only protons are superfluid.

\section{Conserving approximations for superconductors.}

Let us remind first the conserving approximation scheme for normal systems as developed 
in Refs.~\cite{BaymKad,Baym}. 
Since we are looking for collective excitations corresponding to density fluctuations, 
we are interested in the vertex function $\Lambda$, defined by
\beq
\Lambda(12;3)\, =\, <T(\rho\, '(3)\psi^\dagger(1)\psi(2)>
\label{eq:Lambda}
\eeq
where each symbol $i \equiv ({\bf r}_i,\sigma_i,\tau_i,t_i)$ 
stands for coordinate, spin, isospin and time variables of a single particle state; 
$\psi^\dagger(i) , \psi(i)$ are the creation and annihilation operators
for the considered particles.
For given spin and isospin values, the particle density is defined as
$\rho(i) \,=\, \psi^\dagger(i)\psi(i)$
and $\rho\, '(i) \,=\, \rho(i) \,-\, <\rho(i)>$ is the density fluctuation.
Following the standard conventions, 
the single particle Green's function $G(12)$ is defined as
\beq G(12)\,=\, -i<T\{\psi(1)\psi^\dagger(2)\}> \eeq
It satisfies the Dyson's equation
\be 
G^{-1}(12)&=& G_0^{-1}(12)-U(12)-\Sigma(12)
\label{eq:Dyson}
\ee
where $\Sigma(12)$ is the self-energy and $G_0$ the Green's function for non-interacting particles. We
have also introduced a possible external single particle potential $U$, eventually local in space and time and
acting on the particle density. The vertex function $\Lambda$ of Eq.~(\ref{eq:Lambda}) describes the linear
response of the system to such a local external potential
\beq
\Lambda(12;3) \,=\, i\frac{\delta G(12)}{\delta U(3)}
\eeq
where the functional derivative is taken at $U = 0$. By using the chain properties of the functional
derivative, one can derive the integral equation for the vertex function
\be
\Lambda(12;3)&=& \Lambda_0(12;3) +\Lambda_0(12;\overline{1^{\prime}}\overline{2^{\prime}})
\mV(\overline{1^{\prime}}\overline{2^{\prime}};\overline{4}\overline{5}) \Lambda(\overline{4}\overline{5};3)
\label{eq:RPA}
\ee
where a bar over a symbol $i$ indicates integration and summation over the corresponding set of
variables, and
\be 
\Lambda_0(12;3)&=&\frac{1}{i} G(13) G(32)
\label{eq:GG} 
\ee
The effective and irreducible interaction $\mV$ can be expressed in terms of the functional derivative
\be 
\mV(12;45)&=&i\frac{\delta\Sigma(12)}{\delta G(45)}
\label{eq:veff}
\ee
here the functional derivative is meant performed considering the {bf Green's} function as a functional of
the external potential.

If the self-energy $\Sigma$ is approximated with a functional of the Green's function and of the interaction,
both Green's function and self-energy must be calculated by self-consistent procedure. This is the first condition
that a conserving approximation must fulfil. In addition to the self-consistency, one has to impose a symmetry
condition on the self-energy, that is automatically satisfied if it can be obtained as a functional derivative
\beq
\Sigma(12) \,=\, \frac{\delta\, \Phi}{\delta G(12)}
\label{eq:phi}
\eeq
Here $\Phi$ is formally a functional of the Green's function and of the interaction. 
Approximations of this type for the self-energy are called "$\Phi$-derivable" approximations
and are automatically conserving~\cite{Baym}.
It follows from Eqs.~(\ref{eq:Dyson},\ref{eq:RPA},\ref{eq:veff}) that a conserving approximation
for the Green's function entails a conserving approximation for the 
vertex function $\Lambda$, provided
the irreducible interaction $\mV$ is obtained from Eq.~(\ref{eq:veff}). 

To make clear what a conserving approximation actually means, let us remind that the average density $<n(\rv,t)>$
and current $<\rj(\rv,t)>$ can be obtained from the single particle Green's function
\be 
<n(\rv,t)>\, &=&\, \Sigma_{\sigma}\, <\psi^\dagger (\rv,\sigma, t)\psi(\rv, \sigma, t)>
\, =\,i\Sigma_{\sigma} G(x,x^+) \\ \nonumber
<\rj(\rv,t)>\, &=&\, i \Sigma_{\sigma}\, {\hbar \over{2m}}(\nabla_\rv - \nabla_{\rv'})
<\psi^\dagger (\rv, \sigma, t)\psi(\rv', \sigma, t)>|_{\rv' = \rv}
\, =\, \Sigma_{\sigma} (\nabla_\rv - \nabla_{\rv'})G(x,x^+)|_{\rv' = \rv}
\label{eq:current}\ee
where $x^+ = (\rv, \sigma, t+\epsilon)$ (the isospin variable has been neglected), 
with $\epsilon$ a positive infinitesimal quantity that
enables to fix the correct time ordering of the operators in the equal time limit, 
according to the Green's function definition. 
A conserving approximation is such that the so obtained density and current satisfy the local
conservation law
\beq
{\partial \over { \partial t }} <n(\rv,t)> \, +\, \nabla <\rj(\rv,t)> \, =\, 0 
\label{eq:cons}
\eeq

The main reason of this result is the fact that, under the $\Phi$-derivable conditions, the contribution of
the interaction term to the divergence of the current vanishes, and the conservation law of Eq.~(\ref{eq:cons})
then follows trivially from the kinetic term contribution. It has to be noticed that the interaction must be
local, otherwise the conservation law cannot be written in such a local form. The Hartree and Hartree-Fock
approximations for the self-energy are noticeable examples of conserving and $\Phi$-derivable approximations.
These approximations generate the so called Random-Phase-Approximation (RPA) for the vertex function in the
integral equation (\ref{eq:RPA}). Then this approximation for $\Lambda$ conserves the current.

The conserving approximation scheme was originally developed~\cite{Baym} 
for a normal system and for central particle-particle interaction. 
The theory can be readily generalized to a more complex interaction and 
to the case of superfluid systems.

First we have to extend the general formalism to the superfluid case.  
The inclusion of superfluidity can be formally achieved by including in the above scheme 
a further discrete variable $\alpha$ labeling the destruction
($\alpha = 1$) and creation ($\alpha = -1$) operators, 
so that now the collective index 
is $i \equiv ({\bf r}_i,\sigma_i,\tau_i,t_i,\alpha_i) \equiv (x_i,\alpha_i)$. 
Some care must be used in generalizing several
relationships from the normal to the superfluid case. 
In particular, the equation defining the inverse Green's
function $G(12)^{-1}$ must be written
\beq 
\int_{x_3, \alpha_3} G^{-1}(1; x_3,\alpha_3) G(x_3,-\alpha_3;2) \,= \, \delta(1 - 2)
\label{eq:inv} 
\eeq
Then the single particle self-energy $\Sigma(12)$
contains a normal $\Sigma^{n}$ and anomalous part
$\Sigma^{a}$, corresponding to $\alpha_{1}\, \neq\, \alpha_{2}$ and $\alpha_{1}\, =\, \alpha_{2}$, respectively,
and the Dyson's equation (\ref{eq:Dyson}) still holds including the additional variable. Similarly, the
generalized vertex function can be introduced by functional differentiation
\beq
\Lambda(12;x_3,\alpha_3,\alpha_3') 
\,=\, i\frac{\delta G(12)}{\delta U(x_3,\alpha_3,\alpha_3')} 
\,=\, <T(\psi(1)\psi(2)n\,'(x_3,\alpha_3,\alpha_3')>
\eeq
where now $n(x_3,\alpha_3,\alpha_3') \,=\, \psi(x_3,\alpha_3)\psi(x_3,\alpha_3')$ 
is the generalized density, $n' \,=\, n \,-\, <n>$ the corresponding fluctuation, 
and $U$ couples linearly to $n$. 
Using again the chain property of the functional derivative 
one gets the integral equation for the generalized vertex function
\be 
\Lambda(12;3)&=& \Lambda_0(12;3) +\Lambda_0(12;\overline{1^{\prime}}\overline{2^{\prime}})
\mV(\overline{1^{\prime}_{\star}}\overline{2^{\prime}_{\star}};\overline{4}\overline{5})
\Lambda(\overline{4}\overline{5};3)
\label{eq:GRPA} 
\ee
where a subscript $\star$ indicates that a sign change of the variable $\alpha$ has to be considered,
e.g. $1_{\star} = (\rv_1,\sigma_1,t_1, -\alpha_1) $. The function $\Lambda_0$ has the same expression as in 
Eq.~(\ref{eq:GG}), provided the {bf Green's} functions are generalized to the superfluid case. The correlation function
$\Lambda(12;3)$ and the integral equation (\ref{eq:GRPA}) are the generalization to the superfluid 
case of Eqs.~(\ref{eq:Lambda}) and (\ref{eq:RPA}), respectively. Also within this extension of the formalism, an approximation
is conserving if it is $\Phi$-derivable and the Green's function is calculated in a self-consistent way. In the
mean field approximation, which generates the (generalized) RPA approximation for the vertex function, this means
that the self-energy appearing in the RPA equation for the response function must be calculated self-consistently
within the mean field approximation. For the pairing part this means that we have to use the BCS {bf Green's}
functions.

To illustrate this conserving approximation, we will check the conservation of current in the case
of a pure pairing interaction . Let us take a zero-range interaction, that is the one we are going to use in this
work
\beq \hat{U}_{\rm pair} \, =\, -{1\over 2}\,\,\, U_{\rm pair}\,\,\, \Sigma_{\sigma,\sigma'}\,\, \int d^3 r
\psi^\dagger(\rv,\sigma)\psi^\dagger(\rv,-\sigma) \psi (\rv,-\sigma')\psi (\rv,\sigma')(-1)^{(1-\sigma-\sigma')}
\label{eq:int} \eeq
 \noindent
where spin-zero pairing has been assumed and $U_0 $ is the corresponding strength. 
The relevant component of the Green's function $G(11')$ 
corresponds to $\alpha_1 = 1$ and $\alpha_1' = -1$.  This component will be indicated
by $G(x_1,x_1')$. The equation of motion for $G$ can be written
\be
\label{eq:eqm1}
i\hbar{\partial\over {\partial t}} G(x_1,x_1') + {\hbar^2\over {2 m}}\nabla_r^2 G(x_1,x_1') 
&\, =\,& \delta(t - t')\delta(\rv - \rv') \\
&+& i U_{\rm pair}\Sigma_{\sigma'}\,\, 
<T(\psi^\dagger(\rv, -\sigma_1,t)\psi(\rv,-\sigma', t)\psi(\rv, \sigma', t)
\psi^\dagger(\rv', \sigma_1, t'))> (-1)^{(1-\sigma_1-\sigma')} \nonumber
\ee
where $x_1 = (\rv,\sigma_1,t)$ and $x_1' = (\rv',\sigma_1,t')$. 
The analogous equation for the derivative on the time $t'$ reads
\be 
\label{eq:eqm2}
i\hbar{\partial\over {\partial t'}} G(x_1,x_1') - {\hbar^2\over {2 m}}\nabla_r^2 G(x_1,x_1') 
&\, =\,& \delta(t - t')\delta(\rv - \rv')  \\ \nonumber
&+& i U_{\rm pair} \Sigma_{\sigma}\,\, 
<T(\psi(\rv, \sigma_1, t)\psi^\dagger(\rv', \sigma, t')\psi^\dagger(\rv', -\sigma, t')
\psi(\rv', -\sigma_1, t'))> (-1)^{(1+\sigma_1-\sigma)}
\ee
Summing up the two equations of motion, performing the summation over the spin $\sigma_1$, 
and finally putting $t = t'$ and $\rv = \rv'$, 
the first two terms on the right hand side give the conservation law~(Eq.~(\ref{eq:cons})),
because they correspond to the kinetic term contribution. 
The interaction contribution on the right hand side,
that involves the two-body Green's function $G_2$, factorizes in the mean field approximation. 
For the $G_2$ contribution in Eq.~(\ref{eq:eqm1}) one has:
\be
<T(\psi^\dagger(\rv,-\sigma_1,t)\psi(\rv,-\sigma',t)
\psi(\rv,\sigma',t) \psi^\dagger(\rv',\sigma_1,t')> 
&\approx&
<T(\psi^\dagger(\rv,-\sigma_1,t) \psi^\dagger(\rv',\sigma_1,t'))> <\psi(\rv,-\sigma',t)\psi(\rv,\sigma',t)>
\nonumber \\
& + &
<\psi^\dagger(\rv,-\sigma_1,t)\psi(\rv,-\sigma',t)> <T(\psi(\rv,\sigma',t)\psi^\dagger(\rv',\sigma_1,t'))>
\nonumber \\
& - &
<\psi^\dagger(\rv,-\sigma_1,t)\psi(\rv,\sigma',t)> <T(\psi(\rv,-\sigma',t)\psi^\dagger(\rv',\sigma_1,t')>
\nonumber \\
\ee
A completely analogous factorization applies
to the $G_2$ contribution in Eq.~(\ref{eq:eqm2}). 
Taking into account the translational and time reversal symmetries
\beq 
<T(\psi^\dagger(\rv,-\sigma_1,t)\psi^\dagger(\rv',\sigma_1,t')>
\, =\, -<T(\psi(\rv,-\sigma_1,t)\psi(\rv',\sigma_1,t')>
\eeq
the two factorized $G_2$ are equal, but they are multiplied by an opposite phase in the equations of
motion and therefore they cancel out. This conservation property is translated to the vertex function if the
effective interaction $\mV$ is calculated by the functional derivative of equation (\ref{eq:veff}), 
properly generalized to include superconductivity.

This property of the RPA equation, when only pairing interaction appears, 
was recognized a long time ago~\cite{Schriefferb}. 
If other interactions are present, besides the pairing one, noticeably the Coulomb interaction,
the argument can be repeated and this RPA property still holds, provided the additional interaction is local. 
The mean field correction to the self-energy in the mean field approximation amounts to a shift in the chemical potential, 
if the exchange term is neglected. The latter approximation is equivalent to the Hartree
approximation, which is also conserving. If the interaction is non-local, or we include the exchange term, the
local continuity equation (\ref{eq:cons}) does not hold any-more. It can be restored in the effective mass
approximation, where the effective mass is momentum and energy independent. 

It has to be noticed that the conservation law of the vertex function $\Lambda$, 
appearing in Eq.~(\ref{eq:GRPA}), is equivalent to the
generalized Ward identities~\cite{Schriefferb}.

A direct consequence of the conservation law is the behavior of the density-current polarization tensor, 
defined as
\beq
\Pi_{\nu \mu}(x,x') \,=\, <T(j_\nu(x) j_\mu(x'))>
\label{eq:Pi}
\eeq
where $\nu, \mu \,=\, 0,1,2,3$ and $j_\nu$ is the current four vector, 
with $j_0(x) \,=\, \rho(x)$ the particle density. 
This is the response function to a local external probe, which is linearly coupled to $j_\mu$,
provided the particle-hole interaction is local and the pairing interaction is zero range. 
It can be written in terms of the vertex function in a straightforward way~\cite{Schriefferb}. 
The conservation law for $\Pi$ in the energy-momentum ($\omega$,$\qv$) representation can be written
\beq
\omega\,\, \Pi_{(0 \mu)} \, +\, \Sigma_{i=1,3}\ \  q_i\,\, \Pi_{(i \mu)} \, =\, 0
\eeq
For an isotropic system the second (current) term is proportional to $\qv^2$ for small $|\qv|$, and thus
for any value of the energy $\omega$ the density-density component of the polarization tensor, in the limit of
small $|\qv|$, vanishes at least as $\qv^2$.

To conclude this section, we will write explicitly the self-energy in the mean field approximation and the
corresponding effective interaction for the RPA equation. Following the factorization of the two-body Green's
function $G_2$ illustrated above, one can extract the self-energy for a general interaction $v$
\be \Sigma(12)=\frac{i\delta(t_2-t_1)}{2}\int_{x_{i^{\prime}}} &&\left[
2\delta_{\alpha_2,-\alpha_1}<x_1x_{1^{\prime}}|v|x_2x_{2^{\prime}}>_A \left(
\mbG(x_{2^{\prime}},t_1^{+},-1;x_{1^{\prime}},t_1^{-},1)\delta_{\alpha_1,-1} -
\mbG(x_{1^{\prime}},t_1^{+},-1;x_{2^{\prime}},t_1^{-},1)\delta_{\alpha_1,1}\right)
\right.\nonumber\\
&&\left. +\delta_{\alpha_2,\alpha_1}<x_1x_2|v|x_{1^{\prime}}x_{2^{\prime}}>_A
\mbG(x_{1^{\prime}},t_1^{+},\alpha_1;x_{2^{\prime}},t_1^{-},\alpha_1)\right] \label{Eq:GSigma} 
\ee
The effective interaction $\mathcal{V}$, to be used in RPA equations, can be derived
from Eq.~(\ref{Eq:GSigma}):
\be
\mathcal{V}(12;45) =\frac{\delta\Sigma(12)}{\delta\mbG(45)} &=&i\delta(t_2-t_1)\delta_{\alpha_2,-\alpha_1}
\delta_{\alpha_4,-1}\delta_{\alpha_5,1} \left[ <x_1x_5|v|x_2x_4>_A\delta_{\alpha_1,-1} -
<x_1x_4|v|x_2x_5>_A\delta_{\alpha_1,1} \right]
\nonumber\\
&+& \frac{i}{2}\delta(t_2-t_1)\delta_{\alpha_2,\alpha_1} \delta_{\alpha_4,\alpha_1}\delta_{\alpha_5,\alpha_1}
<x_1x_2|v|x_4x_5>_A
\label{eq:GVeff}
\ee
Again, if we assume that the interaction is the sum of a local particle-hole interaction and a local zero-range
pairing interaction, the expression simplifies, and, as we will see, the energy-momentum representation is
particularly convenient.

\section{Equations for the response function}

In the BCS framework, the single-particle Green's function reads:
\be \mbG^{12}(k)&=& \alpha_1\left[\delta_{\alpha_1,-\alpha_2}\delta_{\sigma_1,\sigma_2}G(k)
+\sigma_1\delta_{\alpha_1,\alpha_2}\delta_{\sigma_1,-\sigma_2}F(k)\right] \een
with
\be
G(k)&=&\frac{v_k^2}{E_k+\omega-i\eta}-\frac{u_k^2}{E_k-\omega-i\eta}\nn\\
F(k)&=&u_kv_k\left[\frac{1}{E_k-\omega-i\eta}+\frac{1}{E_k+\omega-i\eta}\right]\nn\\
u_k^2&=&\frac{1}{2}\left(1+\frac{\epsilon_k-\mu}{E_k}\right)\nn\\
v_k^2&=&\frac{1}{2}\left(1-\frac{\epsilon_k-\mu}{E_k}\right)\nn\\
\epsilon_k&=&\frac{\hbar^2k^2}{2M}\nn\\
E_k&=&\sqrt{(\epsilon_k-\mu)^2+\Delta^2} 
\een

For simplicity we assume the pairing to be in the s-wave.
We also consider the pairing interaction to be constant with a cutoff in momentum space,
which corresponds to a contact interaction in coordinate representation.
Furthermore, the interaction in particle-hole channel is local
(it is a central density-density Coulomb interaction).
In such conditions, the general equation (\ref{eq:GRPA}) for the vertex function reduces 
to an equation directly for the polarization function ${\Pi}({\bf q},\omega)$, 
in its generalized form, appropriate to superconductors.
As usual, the momentum representation is obtained by Fourier transform.
From Eq.~(\ref{eq:GRPA}) one gets a system of four coupled algebraic equations to be solved
for $\Pi({\bf q},\omega)$; however it turns out that one equation is a linear combination of the others and the
coupled components of the polarization tensor to be calculated are reduced to three.
Finally, the generalized RPA equations read:
\be 
\left(\begin{array}{ccc}
1-X^{pp}_{-}U_{\rm pair} &  -X_{GGC}U_{\rm pair} & -2X_{GF}^{+}v_c \\[0.25cm]
-X_{GGC}U_{\rm pair} & 1-X^{pp}_{+}U_{\rm pair} & -2X_{GF}^{-}v_c \\[0.25cm]
X_{GF}^{+}U_{\rm pair} & X_{GF}^{-}U_{\rm pair} & 1-2X^{ph}_{-}v_c \\[0.25cm]
\end{array}\right)
\left(\begin{array}{c}
\Pi^{(-)}_S \\[0.25cm]
\Pi^{(+)}_S \\[0.25cm]
\Pi^{(ph)}_S\\[0.25cm]
\end{array}\right)
&=& \left(\begin{array}{c}
\Pi^{(-)}_{0,S} \\[0.25cm]
\Pi^{(+)}_{0,S} \\[0.25cm]
\Pi^{(ph)}_{0,S} \\[0.25cm]
\end{array}\right)
\label{eq:RPA1} 
\ee
where the index $S$ specifies that scalar (zero total spin) excitations are considered, and
we have introduced the notation:
\be
X_{\pm}^{pp}&=&\frac{1}{2}\left[X_{GG}^{pp}(q)+X_{GG}^{pp}(-q)\right] \pm X_{FF}(q) \label{eq:Xpp}\\
X_{\pm}^{ph}&=&X_{GG}^{ph}(q) \pm X_{FF}(q) \label{eq:Xph} \\
X_{GF}^{\pm}&=&X_{GF}(q)\pm X_{GF}(-q) \label{eq:Xgf} \\
X_{GGC}&=&\frac{1}{2}\left[X_{GG}^{pp}(-q)-X_{GG}^{pp}(q)\right]  
\ee
The quantities $X$ are the following four-dimensional integrals
\be 
X_{GG}^{ph}(q)&=&\frac{1}{i}\int\frac{dk}{(2\pi)^4}G(k)G(k+q)\;\;;\;\;
X_{GG}^{ph}(-q)=X_{GG}^{ph}(q) \\
X_{GG}^{pp}(q)&=&\frac{1}{i}\int\frac{dk}{(2\pi)^4}G(k)G(-k+q) \\
X_{GG}^{pp}(-q)&=&\frac{1}{i}\int\frac{dk}{(2\pi)^4}G(k)G(-k-q) \\
X_{GF}(q)&=&\frac{1}{i}\int\frac{dk}{(2\pi)^4}G(k)F(k+q) \\
X_{GF}(-q)&=&\frac{1}{i}\int\frac{dk}{(2\pi)^4}G(k)F(k-q) \\
X_{FF}(q)&=&\frac{1}{i}\int\frac{dk}{(2\pi)^4}F(k)F(k+q)\;\;;\;\; X_{FF}(-q)=X_{FF}(q) 
\ee
They can be considered the different components of 
a generalized Lindhard function, suitable for the study of superfluid systems. 
Their explicit expressions are given in the Appendix. 
If the alpha indices are explicitly indicated, the polarization tensor can be written
\beq 
\mathbf{\Pi}({\bf q},\omega)\, =\, \Pi_{\alpha_1 \alpha_1' ; \alpha_2 \alpha_2'}({\bf q},\omega) 
\eeq
and the three components of the polarization tensor appearing in Eq.~(\ref{eq:RPA1}) are defined as
\be
\Pi^{(+)} &=&\frac{1}{2}\left(\Pi_{11;\alpha\beta}+\Pi_{-1-1;\alpha\beta}\right)  \\
\Pi^{(-)} &=&\frac{1}{2}\left(\Pi_{11;\alpha\beta}-\Pi_{11;\alpha\beta}\right)    \\
\Pi^{(ph)}&=& \Pi_{-1 1; \alpha\beta}
\label{eq:Pimix}
\ee
where the values of the variables $\alpha$ and $\beta$ are generic and are chosen according to the
polarization tensor components that have to be calculated.

In the so-called constant level density approximation, i.e. exact particle-hole symmetry, the off-diagonal
matrix elements of Eq.~(\ref{eq:RPA1}), coupling the first equation with the other two, are of order $q^2$.
As a result, to order $q^4$, the equation involving the component $\Pi^{-}_S$ is decoupled from the other two,
and actually can be neglected.

In the case of neutron star matter, one must take into account the coupling of protons and electrons by the
Coulomb interaction. In this case the complete RPA equations form a system of three coupled equations for the
polarization tensor, involving both proton and electron components
\be 
\left(\begin{array}{ccc}
1-X^{pp}_{+}U_{\rm pair} & -2X_{GF}^{-}v_c & 2X_{GF}^{-}v_c \\[0.25cm]
X_{GF}^{-}U_{\rm pair} & 1-2X^{ph}_{-}v_c  & 2X^{ph}_{-}v_c \\[0.25cm]
      0            &  2X^{e}v_c & 1-2X^{e}v_c \\[0.25cm]
\end{array}\right)
\left(\begin{array}{c}
\Pi^{ (+)}_S \\[0.25cm]
\Pi^{(ph)}_S \\[0.25cm]
\Pi^{(ee)}_S\\[0.25cm]
\end{array}\right)
&=& \left(\begin{array}{c}
\Pi^{(+)}_{0,S} \\[0.25cm]
\Pi^{(ph)}_{0,S} \\[0.25cm]
\Pi^{(ee)}_{0,S} \\[0.25cm]
\end{array}\right)
\label{eq:RPA2} 
\ee
where $X^e$ is the relativistic Lindhard function for electrons, that will be calculated in the Vlasov
limit, and $\Pi^{(ee)}$ is the corresponding part 
of the polarization tensor involving the electron (density) component.

\section{The excitation spectrum}

The simplest situation is when only the pairing interaction is included. 
For the considered case of Eq.~(\ref{eq:RPA2}) this is formally equivalent 
to put $v_c = 0$, and it corresponds e.g. to an uncharged superfluid. 
The corresponding expression for $\Pi^{(ph)}_S$ can be easily obtained by solving the remaining two
by two algebraic system. In the notation introduced in the previous section, 
and using the expressions of $\Pi^{(+)}_{0,S}$ and $\Pi^{(ph)}_{0,S}$ given in Appendix,
it reads
\be 
\Pi^{(ph)}_S \,=\, 2\left[ (1-X^{pp}_{+}U_{\rm pair})X^{ph}_{-} -  (X_{GF}^{-})^2U_{\rm pair}
   \right]/(1-X^{pp}_{+}U_{\rm pair})
\label{eq:opair} 
\ee
One can verify that the numerator of this expression vanishes at $\qv = 0$ for any non-zero value of
$\omega$, which implies that the response function , and the corresponding strength function, are proportional to
$q^2$ for small q for all non-zero value of $\omega$. As shown in the previous section, this is a consequence of
the conserving approximation we are following, which guarantees the conservation of current. Indeed, the
continuity equation implies this property of the density-density component of the polarization tensor. This RPA
property of current conservation is a well known result~\cite{Schriefferb} in the theory of superconductors,
where only pairing interaction is present, and it is also a consequence of gauge invariance.

According to a general theorem on broken symmetry, the strength function for $\omega < 2\Delta$ is characterized
by the presence of a "Goldstone boson" , i.e. a phonon-like excitation with an energy proportional to $|\qv|$ for
small enough $|\qv|$. It is well known~\cite{Schriefferb} that in this limit the velocity of this mode is $v_F/
\sqrt{3}$, provided the weak coupling approximation is valid, where $v_F$ is the Fermi velocity. The energy of the
Goldstone mode is obtained just by the equation
\be  
1-X^{pp}_{+}U_{\rm pair}\, =\, 0 
\label{gold}
\ee
which corresponds to the vanishing of the determinant of the matrix 
on the left hand side of Eq.~(\ref{eq:RPA2}). As expected, according to Eq.~(\ref{eq:opair}), 
the strength function has a delta singularity at the energy of the mode, 
since it is undamped in the considered limit of pairing interaction only.

It is a classical result~\cite{Schriefferpr} that in a charged superconductor the Goldstone mode cannot be present
since the standard theorem on broken symmetry (the gauge invariance in the present case) is not any more valid if
the particles interact also by a long range interaction (the Coulomb potential). The mode is replaced by the
plasmon mode. The plasmon excitation has an energy which starts at a non zero value at $\qv = 0$ (the plasmon
energy) and varies slowly with the momentum. This conclusion can be obtained by considering the first two
equations (\ref{eq:RPA2}), including $v_c$ and neglecting the third equation and the electron component. We deal
then with a charged superconductor. One finds that the determinant of the matrix 
on the left hand side of Eq.~(\ref{eq:RPA2}) in this limit vanishes at non-zero energy for $\qv = 0$, since the factor in front of $v_c$
actually vanishes, as it happens in a normal charged fluid.

Here we focus on the case when the pairing gap is comparable to or larger than the plasmon frequency, as it can
happen in high T$_c$ superconductor. It can be true also in neutron star matter for the proton component. 
Even if in the latter case the electron screening changes the physical situation, 
we will discuss the neutron star proton superconductivity, 
stressing the possible analogy with the high T$_c$ superconductors, where the physical
situation is more complex due to the crystal electronic structure. 
Then we will introduce the electron screening
and discuss the effect of proton superfluidity on the electron plasmon.

\begin{figure}[h]
\vskip -8 cm
\begin{center}
\includegraphics[bb= 200 50 300 600,angle=0,scale=0.8]{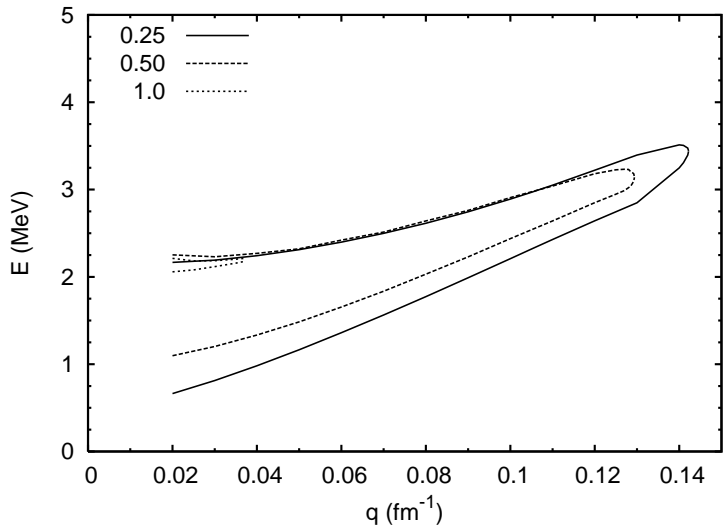}
\caption{Proton excitation spectrum at different values of the pairing gap, as indicated by the labels. Electron
screening is not included.} 
\label{fig:pplasm}
\end{center}
\end{figure}

The first main point that we want to address in a general framework 
is where the position of the plasmon excitation can be expected at increasing value
of the pairing gap. At first sight one could expect that the original excitation spectrum, present in the normal
system, is just shifted upwards by an amount comparable with the energy gap in the superfluid system. However it
turns out that this is not the case.
Let us consider the neutron star matter at a total baryon density equal to the saturation one, i.e. 0.16 fm$^{-3}$. At
this density the proton fraction is expected to be 3.7\%~\cite{paper1,paper2}, and the proton plasmon energy
$\omega_p \,\approx\,$ 2.105 MeV. Under these physical conditions, the full structure of the excitation spectrum
is illustrated in Fig.~\ref{fig:pplasm}, for different values of the pairing gap.

The branches of the excitation spectrum are defined as 
the zeroes of the determinant of the real part of the matrix at the left hand side of Eq.~(\ref{eq:RPA2})
in the energy-momentum space $(\omega,| \qv |)$.
The density and gap values can be considered realistic~\cite{paper1,paper2,hanma}. 
Some comments are in order. 
When the gap $\Delta$ is much smaller than the plasmon energy, the structure of the excitation branches
resembles the "thumb like" shape typical of the charged normal Fermi liquid~\cite{Fetter-Walecka,Jancovici,McOrist,paper1,paper2}. 
In this case the lower branch is over-damped, and indeed
it does not correspond to an actual excitation, while the upper one is just the plasmon mode, that is mainly
undamped up to a momentum nearby the end of the "thumb", above which no excitation mode exists. In the considered
superfluid case, however, the lower branch must start above the forbidden energy region $\omega < 2\Delta$. 
At increasing value of $\Delta$, the spectrum changes considerably. Not only the momentum dependence is modified and
quite different from a quadratic shape, but the whole spectrum is compressed and stays just above $2\Delta$. 
When $2\Delta$ approaches the plasmon energy $\omega_p$ 
of the normal system the spectrum can exist only in a narrow range of momentum 
and finally no excitation is possible. 
Therefore no plasmon excitation is possible inside the forbidden region, 
at variance with high T$_c$ superconductors, 
where a plasmon mode is observed well below 2$\Delta$ along the $c$-axis. 
However in high T$_c$ superconductors the reason of the low value of the plasmon mode is due not only to
the low density of the charge carriers but mainly to the very high value of the insulating dielectric constant
along the $c$-axis~\cite{hightc1,hightc2}. The coherent superfluid motion is possible because the pairing
coherence length is longer than the inter-plane distance in the $c$-axis direction, so that the planes form a
micro-array of Josephson's junctions (hence the name of "Josephson plasmon").

Going back to the neutron star matter, we notice that if the electron component is introduced, the overall
structure of the spectrum is modified. In fact, electrons are much faster than protons and are able to screen the
proton-proton Coulomb interaction at all considered frequencies. Since then the proton-proton effective interaction
is of finite range, a sound-like branch appears again below $2\Delta$. This excitation can be considered a
pseudo-Goldstone mode. It is determined by both pairing and screened Coulomb interaction. 
In fact, without the pairing interaction a sound mode is anyhow present at low momenta~\cite{paper1};
it is strongly damped, and its velocity is
determined by the screened Coulomb interaction. As discussed above, if only the pairing interaction is
considered, a Goldstone mode is necessarily present below $2\Delta$, essentially undamped, with a velocity that in
the weak coupling limit is independent from the interaction strength and equal to $v_F/ \sqrt{3}$. The
pseudo-Goldstone mode turns out to have a much higher velocity, about three times larger.

\begin{figure}[h]
\vskip -8 cm
\begin{center}
\includegraphics[bb= 200 50 300 600,angle=0,scale=0.8]{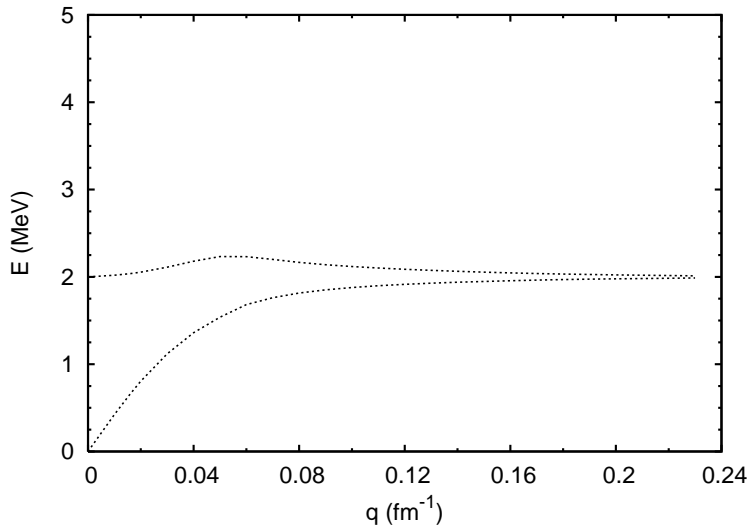}
\caption{The branches of the proton excitation spectrum for $\Delta \,=\, 1$ MeV. Electron screening is included.}
\label{fig:brand10}
\end{center}
\end{figure}

The structure of the excitation spectrum is illustrated in Fig.~\ref{fig:brand10} for $\Delta =$ 1 MeV and at the same total
baryon density. The reported branches correspond to the ones where the proton strength is dominant. One notices
the pseudo-Goldstone mode below 2$\Delta$ and the pair-breaking mode close to 2$\Delta$. At increasing momentum
the pseudo-Goldstone branch deviates from the linear behavior and approaches slowly 2$\Delta$ and the two branches
seem finally to touch at an end point of the spectrum.

For much smaller value of the gap the spectrum looks slightly different. The pseudo-Goldstone mode and the
pair-breaking mode actually cross.
Above the crossing the upper branch corresponds to the sound mode present of
the normal system, while the lower branch corresponds to the over-damped mode that is also present in the normal
system. Roughly speaking, the "thumb-like" shape of the spectrum above 2$\Delta$ resembles the similar shape that
is present in the normal system~\cite{paper1,paper2}. We will not discuss in detail the evolution of the spectrum
as a function of the gap value since this is outside the scope of the present work. In any case, if the value of
the gap is vanishingly small, one recovers the spectral shape of the normal system.

Up to now we have neglected the nuclear interaction. The inclusion of the proton-proton and proton-neutron
effective interactions in the description of the collective modes in superfluid neutron star matter is beyond the
scope of the present work, and we will keep the discussion at a qualitative level. 
In previous papers~\cite{paper1,paper2} we have shown that the neutron-proton coupling in the very asymmetric matter of neutron stars
is weak, and therefore the proton excitations are approximately decoupled from the neutron component. 
The nuclear proton-proton interaction can of course modify the velocity of the pseudo-Goldstone mode, 
but the above scheme of the excitation spectrum should be still valid. 
The detailed study of the general case of superfluid neutron starmatter is left to future works.

\section{The electron plasmon in neutron star matter.}

The spectrum of the electron component in neutron star matter is characterized by a plasmon excitation. Since the
electron Fermi velocity (very close to the speed of light) is much higher than the proton Fermi velocity, the
protons act essentially as a static positive background. However some dynamical electron-proton Coulomb coupling
is still present and produces a relatively small shift of the electron plasmon energy with respect to the purely
static limit. At the plasmon energy some proton strength is then present, but it turns out to be in general quite
small~\cite{paper1,paper2}. If the protons are superfluid another phenomenon occurs, that is not present in the
normal system. The electron plasmon is coupled by the Coulomb interaction to the particle-hole proton excitations,
characterized mainly by a sound mode. If the sound mode energy is higher than 2$\Delta$, it can decay since it is
in turn coupled to the pair-breaking mode by the pairing interaction. This indirect coupling of the electron
plasmon to the proton pair-breaking mode produces a damping of the excitation, and therefore its width increases
at increasing value of the pairing gap. This can be studied by considering the electron strength function
\beq
S_e(\qv,E) \, =\, - \Im (\,\,  \Pi^{(ee)} \,  ) 
\label{eq:stre}
\eeq
where $\Im$ indicates the imaginary part. 
The function $\Pi^{(ee)}$ is the density-density polarization function for the electrons. 
It gives the strength of the electron component at a given excitation energy $E$ and momentum $|\qv|$.

\begin{figure}[h]
\vskip -8 cm
\begin{center}
\includegraphics[bb= 200 50 300 600,angle=0,scale=0.8]{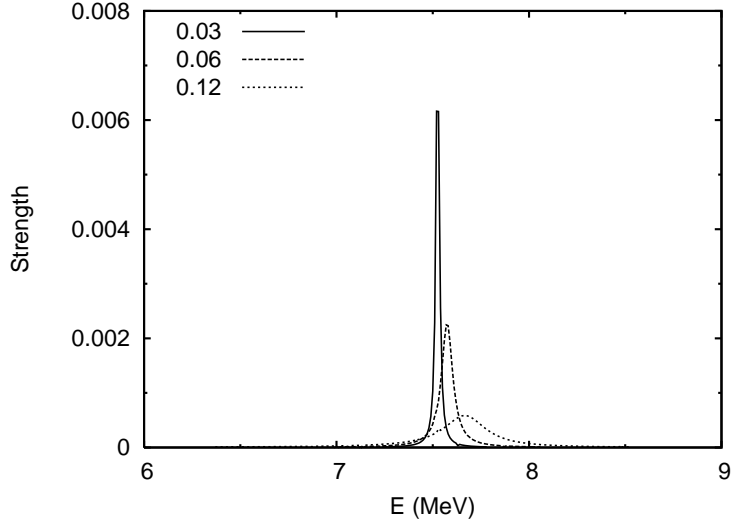}
\caption{Electron strength function at different values of the pairing gap, as indicated by the labels.}
\label{fig:eplasm}
\end{center}
\end{figure}

In Fig.~\ref{fig:eplasm} is reported the electron strength function for $|\qv| = $5 MeV as a function of $E$ and for different
values of the pairing gap, 
again at the total baryon density $\rho \,=\, 0.16$ fm$^{-3}$~\cite{paper1,paper2}.
One can notice the sharp plasmon peak that is obtained for the lowest value of the pairing gap, 
as expected for a normal system. 
However, even for moderately small values of the gap the width of the plasmon increases rapidly and its
strength is spread in a larger and larger energy interval, corresponding to a stronger damping. In fact the
energy-weighted sum rule for the electron strength must be satisfied, as we checked numerically, and therefore the
electron strength is also present at much lower and higher energy with respect to the plasmon energy in the normal
system. This can have some consequence for e.g. the neutrino emission in the cooling process of neutron stars.

\section{Summary and outlook}

We have studied the possible plasmon excitations in strong superconductors, where the pairing gap $\Delta$ is
comparable with the plasmon energy. As a study case we have considered the proton component in the homogeneous
neutron star matter. We found that the plasmon excitation disappears if 2$\Delta$ is equal or larger than the
value of the plasmon energy in the normal system. This should be a general result, not connected with the
particular features of the neutron star matter. This finding is in contrast with what is predicted and observed in
high-T$_c$ superconductor, where a plasmon excitation is predicted and observed well below 2$\Delta$. However, the
physical situation is quite different in these solid state systems, where the smallness of the plasmon energy is
not due to the low density of the charge carriers, but to the high value of the insulating dielectric constant.

In a more realistic description of the neutron star matter one has to include the electron component. This
modifies the spectrum, since the electrons are faster than protons and screen the Coulomb proton-proton
interaction. The plasmon is replaced by a pseudo-Goldstone excitation below 2$\Delta$. Finally, the inclusion of
the particle-hole nuclear effective interaction among protons and neutrons is expected to further modify the
spectrum, but some of the qualitative features of the proton component excitations are expected to remain
valid.

We have then considered the electron plasmon excitation, 
that in neutron star matter occurs at energy much higher than 2$\Delta$.
Although the proton-proton pairing interaction can affect only indirectly the electron excitations, it turns out
that even at moderately small value of the pairing gap the electron plasmon is strongly damped.

On general ground, our results stress the relevance of the Coulomb interaction in strong superconductors, in
particular in neutron star matter.

\appendix

\section{Generalized Lindhard function and numerical method}

The relevant functions $X$ appearing in the kernel of the RPA equations, reported in Eqs.~(\ref{eq:Xpp}-\ref{eq:Xgf}), 
after the integrations over energy, can be written
\be 
X^{pp}_\pm 
&=& 
\frac{1}{2}\int \frac{d^3\kv}{(2\pi)^3} \left(u_\kd u_{\kv+\qv} \pm v_\kd
v_{\kv+\qv}\right)^2 \left[ \frac{1}{E_{\kv+\qv}+E_\kd+q_0-i\eta} +\frac{1}{E_{\kv+\qv}+E_\kd-q_0-i\eta}
\right]\\
X_{-}^{ph}
&=&
- \frac{1}{2}\int \frac{d^3\kv}{(2\pi)^3}
\left(u_\kd v_{\kv+\qv} + v_\kd u_{\kv+\qv}\right)^2
\left[ \frac{1}{E_{\kv+\qv}+E_\kd-q_0-i\eta}+\frac{1}{E_{\kv+\qv}+E_\kd+q_0-i\eta}
\right]\\
X_{GF}^-
&=&
\int\frac{d^3\kv}{(2\pi)^3}\; u_{\kv+\qv}v_{\kv+\qv}
\left[ \frac{1}{E_{\kv+\qv}+E_\kd-q_0-i\eta} -
\frac{1}{E_{\kv+\qv}+E_\kd+q_0-i\eta}
\right]\nn\\
&=&
\int\frac{d^3\kv}{(2\pi)^3}\; u_{\kd}v_{\kd}
\left[ \frac{1}{E_{\kv+\qv}+E_\kd-q_0-i\eta} -
\frac{1}{E_{\kv+\qv}+E_\kd+q_0-i\eta} \right] 
\ee
Despite the approximation of a constant pairing gap, these integrals cannot be done analytically and
their numerical evaluation requires particular care. We follow the method of assigning to the quantity $\eta$ a
small value, simulating an infinitesimal, and then integrating by an adaptive method the remaining
two-dimensional integrals.
We used the subroutine DT20DQ of the Visual Fortran package.
In this way both real and imaginary part can be calculated in few minutes in a simple PC. We found that the
results are quite stable if we take for $\eta$ values between 10$^{-3}$ and 10$^{-6}$ (energies are all
calculated in MeV). For production calculations we used the value 1.5 x 10$^{-5}$.

Finally, for completeness we give here the expression of the free polarization functions appearing on the right
hand side of RPA Eq.~(\ref{eq:RPA2}), in the case we are interested in the density-density response function,
i.e. the spectral function ( $\alpha = -1$ and $\beta =1$ )
\be
\Pi^{(+)}_{0,S} &=& 2 X_{GF}^- \\
 \  \nn\\
\Pi^{(ph)}_{0,S} &=& 2 X_{-}^{ph} \\
 \  \nn\\
\Pi^{(ee)}_{0,S} &=& 0
\ee
Using this expression, for the pure pairing case (i.e. $v_c \rightarrow 0$) one gets the expression of
Eq.~(\ref{eq:opair}) for the density-density component $\Pi^{(ph)}_{S}$. Then, putting $\qv = 0$ in the previous
expressions for the functions $X$, one finds
\beq
U_{\rm pair}\,\, (X_{GF}^-)^2 \, =\, U_{\rm pair}\,\, q_0^2\, \Delta^2 \left[\int\frac{d^3\kv}{(2\pi)^3}\,\,
\frac{1}{E_{\kv}[(2E_\kv)^2 - q_0^2] } \right]^2 \, =\, X_{-}^{ph}\, (1\, -\, U_{\rm pair}\, X^{pp}_+)
\eeq
The relation shows that $\Pi^{(ph)}_{S} (\qv,q_0)$ vanishes at $\qv = 0$ for any value of $q_0$, i.e.
it is proportional to $\qv^2$ in the small $|\qv|$ limit for all $q_0$. This result can be readily generalized
to the case where a finite range particle-hole interaction is present.

\end{document}